\documentclass{PoS}

\title{Improved actions and lattice coarsening effects in MCRG studies in SU(2) LGT}

\ShortTitle{Improved actions and lattice coarsening effects in MCRG studies in SU(2) LGT}

\author{\speaker{E.T. Tomboulis} and Alexander Velytsky\thanks{Current address: 
EFI, University of Chicago, 5640 S. Ellis Ave., Chicago, IL 60637
and Argonne National Laboratory, 9700 Cass Ave., Argonne, IL 60439}\\
        Department of Physics and Astronomy, UCLA, Los Angeles, CA 90095-1547, USA\\
        E-mail: \email{tombouli@physics.ucla.edu},
        \email{vel@physics.ucla.edu}}

\abstract{We study decimation procedures and effective (improved) actions in the framework of Monte Carlo Renormalization Group (MCRG). Particular attention is paid to matching the form of the effective action to the decimation procedure parameters.
Using the static quark-antiquark potential in SU(2) LGT we probe different distance scales and find that an effective action containing multiple group representations is capable of reproducing long distance physics well. In particular, 
appropriate matching results in the practical elimination of the coarsening/fining effect of the lattice spacing under decimation. The short distance regime of the effective theory is also studied. 
We next carry out studies of effective actions involving both multiple representations and loops beyond the single plaquette towards determining an 
improved action good over a wide length scale regime.}

\FullConference{The XXV International Symposium on Lattice Field Theory\\
		 July 30 - August 4 2007\\
		 Regensburg, Germany}

\begin{document}

\section{Decimations and Observables}
We report on our ongoing study of decimation procedures and effective actions 
in the framework of MCRG. It is based on our previous work 
\cite{Tomboulis:2007nn,Tomboulis:2007re,Tomboulis:2007rn} and constitutes a 
natural continuation. The aim is to extend the effective actions 
studied in these previous works so as to obtain a good representation of 
the physics of the original (undecimated) theory over a wide range - 
from short distance (rotational invariance) to long distances
(string tension).

We employ two decimation procedures doubling the lattice spacing 
at various values of parameter $c$ (relative staple weight):
\begin{itemize}
\item Swendsen decimation 
\begin{equation}
Q_\mu(n)=U_\mu(n)U_\mu(n+\hat{\mu})+c\sum_{\nu\neq\mu}U_\nu(n)U_\mu(n+\hat{\nu})
U_\mu(n+\hat{\nu}+\hat{\mu})U_{-\nu}(n+\hat{\nu}+2\hat{\mu}) \label{eq:sdec}
\end{equation}
\item Double Smeared Blocking (DSB)
\begin{eqnarray}
U_\mu(n)&=&(1-6c)U_\mu(n)+c\sum_{\nu\neq\mu}U_\nu(n)U_\mu(n+\hat{\nu})
U^\dagger_\nu(n+\hat\mu)\quad \times 2\, {\rm times}\nonumber\\
Q_\mu(n)&=&U_\mu(n)U_\mu(n+\hat\mu). \label{eq:dsbdec}
\end{eqnarray}
\end{itemize}
First we look at a single plaquette multi-representation effective action:
\begin{equation}
S=\sum_{j=1/2}^{N_r}\beta_j[1-\frac1{d_j}\chi_j(U_p)] \;.\label{eq:ef_act}
\end{equation}
The demon method is used for measurements of couplings. 
For more details of the simulations we refer to the mentioned references.

We compare observables (at different length scales) on the decimated and 
effective-action-generated configurations
\begin{equation}
{\Delta W_{N\times N} \over W_{N\times N}^{dec}}=
\frac{W_{N\times N}^{gen}-W_{N\times N}^{dec}}{W_{N\times N}^{dec}} \;. 
\label{eq:delta}
\end{equation} 
Results are shown in Table \ref{tab:fix_c_dsb}. 
It is important to notice that demon measurements  performed at 
the length scale of a plaquette do not preserve plaquette expectations 
("no pinning"). In fact, as seen from Table \ref{tab:fix_c_dsb}, 
some decimation parameters tend to reproduce better long distance physics 
at the price of distorting short distance observables.

\begin{table}[ht]
 \centering
 \begin{tabular}{|c||c|c|c|c|}
 \hline
 $c$ &$\beta_{1/2}, \beta_1, \beta_{3/2}, \ldots$ & 
 $\Delta W_{2\times2}/ W^{dec}_{2\times2}$ & $\Delta W_{3\times3}/ W^{dec}_{3\times3}$&$\Delta W_{4\times4}/ W^{dec}_{4\times4}$ \\
 \hline \hline
 0.050&2.3536(5),-0.4208(9)&&&\\
      &0.1430(11),-0.0558(13)&-0.1817(6)&-0.637(1)&\\
      &0.0238(13),-0.0094(15)&&&\\\hline
 0.060&2.4660(7),-0.3635(11)&&&\\
      &0.1242(17),-0.0475(21)&0.0105(7)&-0.239(2)&\\
      &0.0195(25),-0.0070(24)&&&\\\hline
 0.063&2.4891(7),-0.3331(11)&&&\\
      &0.1140(14),-0.0436(19)&0.0800(9)&-0.049(3)&\\
      &0.0180(25),-0.0070(25)&&&\\\hline
 0.065&2.5023(7),-0.3098(12)&&&\\
      &0.1057(16), -0.0397(16)&0.1305(9)&0.106(3)&-0.034(14)\\
      &0.0145(14),-0.0029(15)&&&\\\hline
 0.067&2.5125(7),-0.2832(16)&&&\\
      &0.0964(25),-0.0367(29)&0.1774(9)&0.266(3)&0.290(19)\\
      &0.0139(29)&&&\\\hline 
 0.077& 2.5463(11),-0.1167(17),&0.4149(14)&1.270(7)&\\
      & 0.0320(23),-0.0055(28)&&&\\\hline
   0.1&2.4762(20),0.4191(37)&&&\\
      &-0.1231(40),0.0504(39)&0.6558(9)&2.627(6)&\\
      &-0.0191(53),0.0063(54)&&&\\
 \hline
 \end{tabular}
 \caption{\label{tab:fix_c_dsb} DSB decimations. Demon-measured couplings at 
different $c$ values, and
difference of various size Wilson loops measured on decimated
versus effective-action-generated configurations. 
Demon thermalization: 100 sweeps, measurements: 20 sweeps.}
\end{table}
One should indeed note here that there is pronounced dependence on the 
value of the decimation parameter $c$. In particular, different trends as 
a function of scale are seen for different $c$ values. 

Thus, the ability of a given chosen effective action to reproduce the original 
physics in a certain regime can vary substantially with `tuning' of 
the decimation transformation parameters. We refer to this as matching 
the decimation prescription to the assumed form of the effective action.  

\begin{figure}[ht]
  \includegraphics[width=0.9\textwidth]{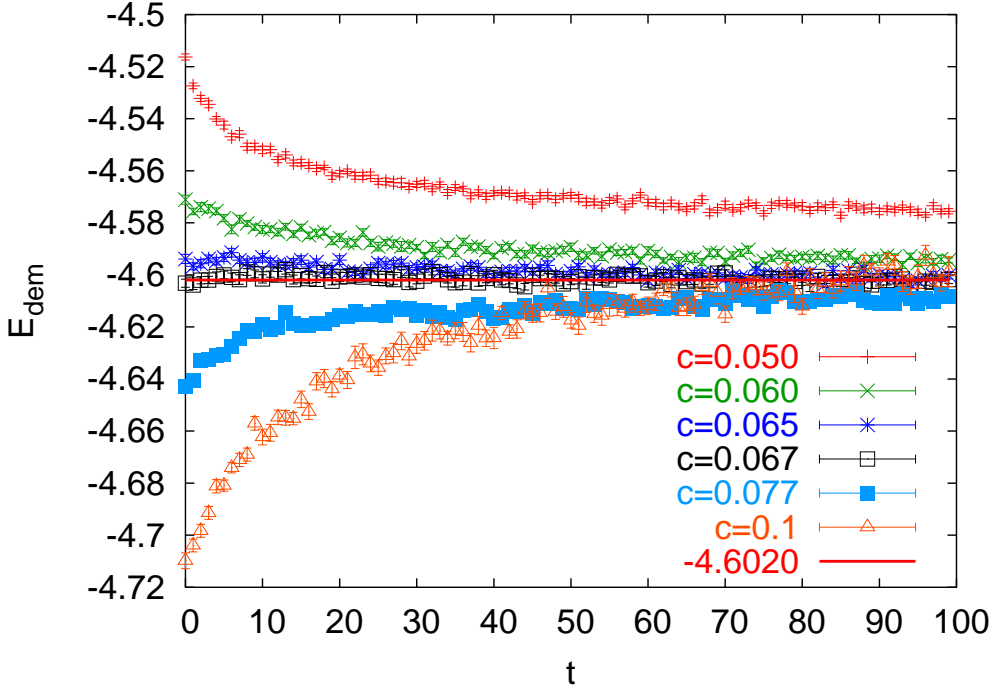}
  \caption{\label{fig:ds_dec_dem}Demon fundamental representation 
energy flow for DSB decimation at various $c$ values.}
\end{figure}

At arbitrary $c$ value there is generally demon energy flow (Fig. 
\ref{fig:ds_dec_dem}), which is a sign
of an on-going thermalization process during the microcanonical evolution. 
This is due to the fact that in 
general the decimated configurations are {\it not} equilibrium 
configurations of the effective action.
Note that the value of $c$ with no demon flow is around the value 
giving better long distance behavior.

\section{Potential and string tension}
\begin{figure}[!ht]
 \includegraphics[width=0.9\columnwidth]{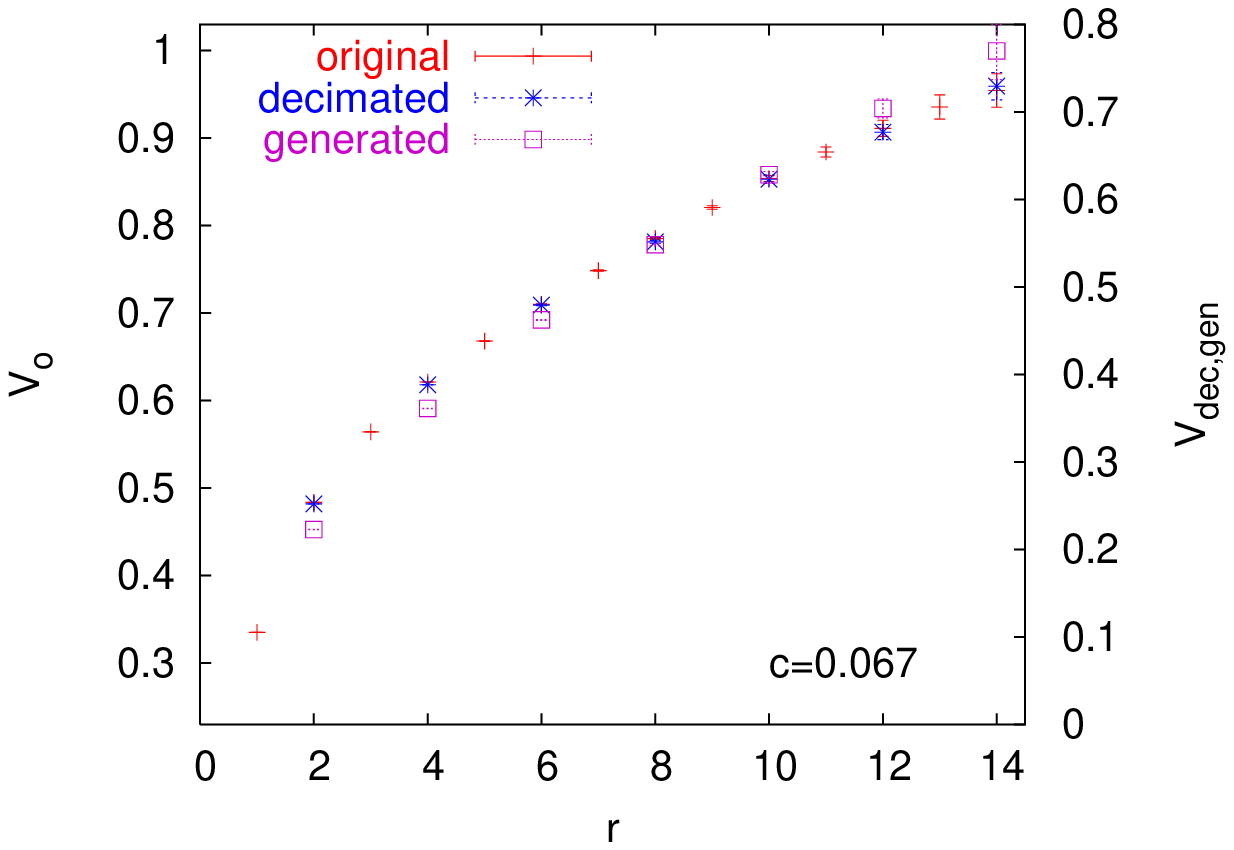}
\includegraphics[width=0.8\columnwidth]{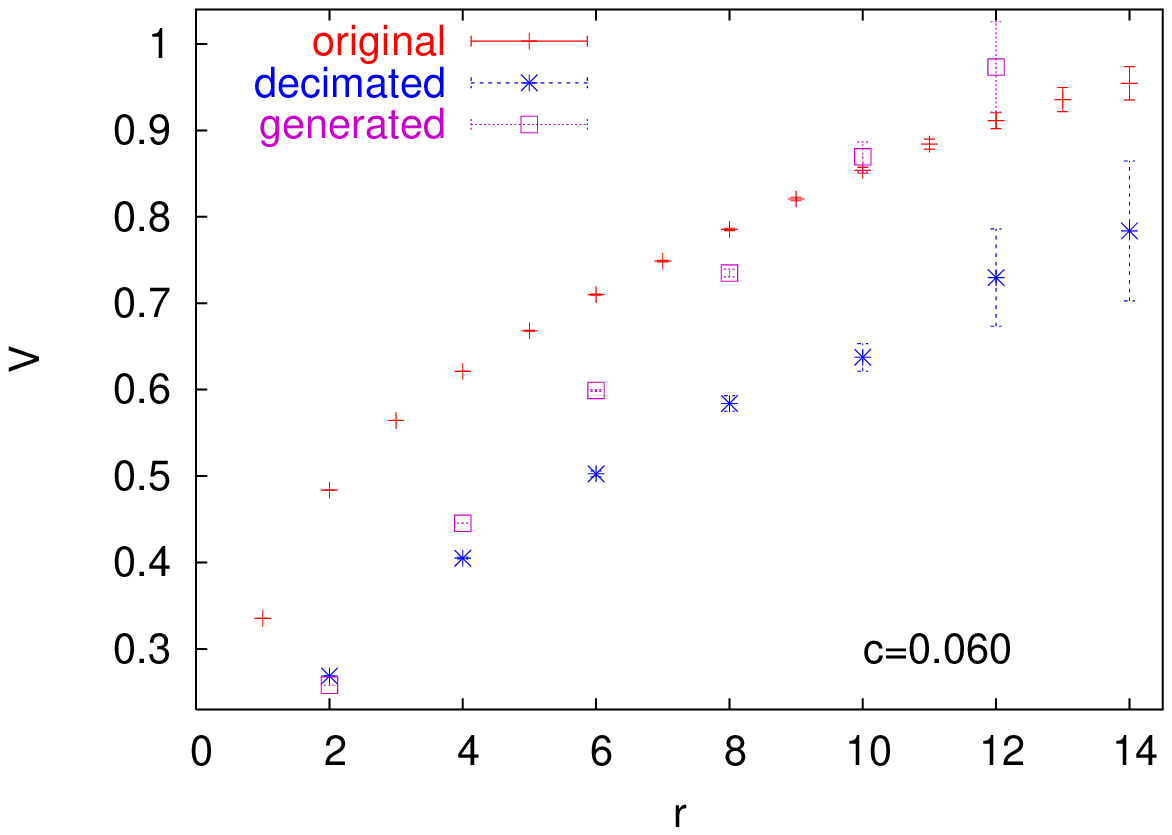}
 \caption{\label{fig:pot1} Static quark potential $V_0$ computed on original 
(undecimated), $V_{dec}$ on DSB decimated ($c=0.067$), and $V_{gen}$ on 
effective-action-generated lattices (top). The same, but for ($c=0.060$)
(bottom).}
\end{figure}

Next, we measure the string tension using the standard ansatz for the 
quark-antiquark potential. It is measured naively for the 
decimated configurations and using 
L\"uscher-Weisz method for the effective-action-generated configurations. 
\begin{table}[ht]
\centering
\begin{tabular}{|c||c|c|}
\hline
   &\multicolumn{2}{|c|}{$\sigma_0=$ 0.0313(2)}\\\hline
c    &$\sigma_{dec}$&$\sigma_{gen}$\\\hline
0.060&0.0271(37)&0.0594(12)\\
0.065&0.0284(30)&0.0385(14)\\
0.067&0.0291(49)&0.0346(8)\\
0.068&0.0295(12)&0.0292(9)\\
0.077&0.0285(24)&0.0091(6)*\\
\hline
\end{tabular}
\caption{\label{tab:string}String tensions in original (undecimated) 
lattice units. $\sigma_0$ on $32^3\times 12$, $\sigma_{dec}$, 
$\sigma_{gen}$ on $16^3\times6$, except 
starred entry which is on $16^3\times 8$.}
\end{table}
Having the string tension allows us to estimate any lattice spacing distortion 
beyond the decimation scaling factor (here equal to $2$) 
resulting from the numerical decimation procedure (so called "coarsening" 
effect).  
Letting 
\begin{equation}
{a_{dec}\over 2a} =\sqrt{\sigma_{dec} \over \sigma_0} \, , \qquad 
{ a_{gen} \over a_{dec}} =\sqrt{\sigma_{gen} \over \sigma_{dec} }\, , \qquad 
{a_{gen}\over 2a} =\sqrt{\sigma_{gen} \over \sigma_0}  \,, \label{eq:ratios}
\end{equation}
the results are shown in Table \ref{tab:ratios}. 
\begin{table}[ht]
\centering
\begin{tabular}{|c||c|c|c|}
\hline 
c    &$a_{dec}/2a $&$ a_{gen}/ a_{dec}$&$ a_{gen} / 2a$ \\\hline
0.060& 0.93(6)&1.48(10)  & 1.378(15)\\
0.065& 0.95(5)&1.16(7) & 1.109(20)\\
0.067&0.96(8)&1.09(9)& 1.051(13)\\
0.068& 0.97(2)&0.99(3)&0.966(15)\\
0.077& 0.95(4)&0.57(3)  & 0.539(18)\\
\hline
\end{tabular}
\caption{\label{tab:ratios} Lattice spacings ratios among original, 
decimated and effective-action-generated lattices.}
\end{table}
One now notes that coarsening is practically absent for values of $c$ close to 
those giving equilibrium values (no demon energy flow). For any other values, 
however, 
very substantial coarsening effects are generated (see Fig. 
\ref{fig:pot1}). In other words, it is imperative to appropriately 
tune the decimation parameter to the effective action under consideration. 

\subsection{Short distance and rotational invariance}
The Wilson action 
on a lattice of spacing $2a$ would have a string
tension $\sigma=(2a/a)^2 \sigma_0$. Therefore, we expect $\sigma\sim0.1252$.
This is very close to the string tension for the Wilson action at 
$\beta=2.31$, which is $\sigma=0.1230(14)$. 
We compare departures from rotational symmetry on this Wilson action lattice
and on the effective-action-generated lattice by looking at the off-axis
correlators.
We quantify the amount of rotational invariance violation by 
\cite{deForcrand:1999bi}:
\begin{equation}
\delta_V^2=\sum_{off}\left(
\frac{V_{off}(r)-V_{on}(r)}{V_{off}(r)\delta V_{off}(r)}\right)^2/
\sum_{off}\frac1{\delta V(r)^2}. \label{eq:deltaV} 
\end{equation}
The results for first four off-axis points are: 
$\delta_V=0.045$ for Wilson, and $\delta_V=0.047$ for the effective 
action. They are comparable - there is no improvement at short distance.

\section{Extended action} 
So far we worked with the multi-representation plaquette effective 
action (\ref{eq:ef_act}). We saw that, appropriately tuning the decimation 
parameter $c$, this action reproduces long-distance behavior very well, but 
shows no particular improvement at short distances.   
In order to achieve such improvement (rotational invariance - removal of
lattice artifacts) while preserving correct long distance behavior we next 
study a multi-representation action including the addition of the $1\times2$ 
loop (rectangle):  
\begin{equation}
S=\sum_{plaq}\sum_{j=1/2}^{j_N}\beta^{plaq}_j[1-\frac1{d_j}\chi_j(U_p)]+
\sum_{rect}\sum_{j=1/2}^{j'_N}\beta^{rect}_j[1-\frac1{d_j}\chi_j(U_{1\times2})]
\label{eq:ef_act_ext} \;.
\end{equation}
We report here some preliminary results with this extended 
action (\ref{eq:ef_act_ext}) in Table \ref{tab:fix_ext_c_dsb}.  
\begin{table}[ht]
\begin{tabular}{|c|l|l|c|}
 \hline
 $c$ &$\beta^{plaq}_{1/2}, \beta^{plaq}_1, \beta^{plaq}_{3/2}, \ldots$ 
	&$\beta_{1/2}^{rect}, \beta^{rect}_1, \beta^{rect}_{3/2}, \ldots$
	&$\sigma_{gen}$\\
 \hline 
\hline
0.90&3.4610(21),-0.0322(28),&-0.3468(4),0.0582(6),&0.0384(34)[12]\\
    &0.0206(33),-0.0105(29) &-0.0221(7),0.0102(7)&\\\hline
1.00&3.4039(19),-0.0357(28), &-0.3262(4),0.0564(6),& 0.0431(31)[22]\\
    &0.0187(35),-0.0112(32)  &-0.0209(6),0.0097(5)&\\\hline
1.40&3.1912(15),-0.0504(21),&-0.2542(4),0.0491(6),& 0.0322(21)[26]\\
    &0.0113(24),-0.0078(20) &-0.0183(7),0.0083(8)&\\\hline
1.50&3.1518(21),-0.0576(22),&-0.2398(4),0.0466(4),& 0.0359(39)\\
    &0.0136(27),-0.0095(20) &-0.0173(6),0.0074(7)&\\
\hline
\end{tabular}
\caption{\label{tab:fix_ext_c_dsb} The couplings of the extended effective 
action after Swendsen decimation starting from the
Wilson action at $\beta=2.5$ and the string tension computed in the corresponding 
effective model. $c$ is the staple weight.}
\end{table}
The $c$ values in Table \ref{tab:fix_ext_c_dsb} are all in the 
range that minimizes demon energies flow. 
We see that the resulting string tension is in the vicinity of the 
correct value. Coarsening 
effects range from 2\% to 10\%. The main systematic
error (not indicated) is due to error in $\beta$ determination.
()  stands for stochastic and [] for systematic (truncation) errors. 

\begin{figure}[ht]
 \includegraphics[width=0.9\columnwidth]{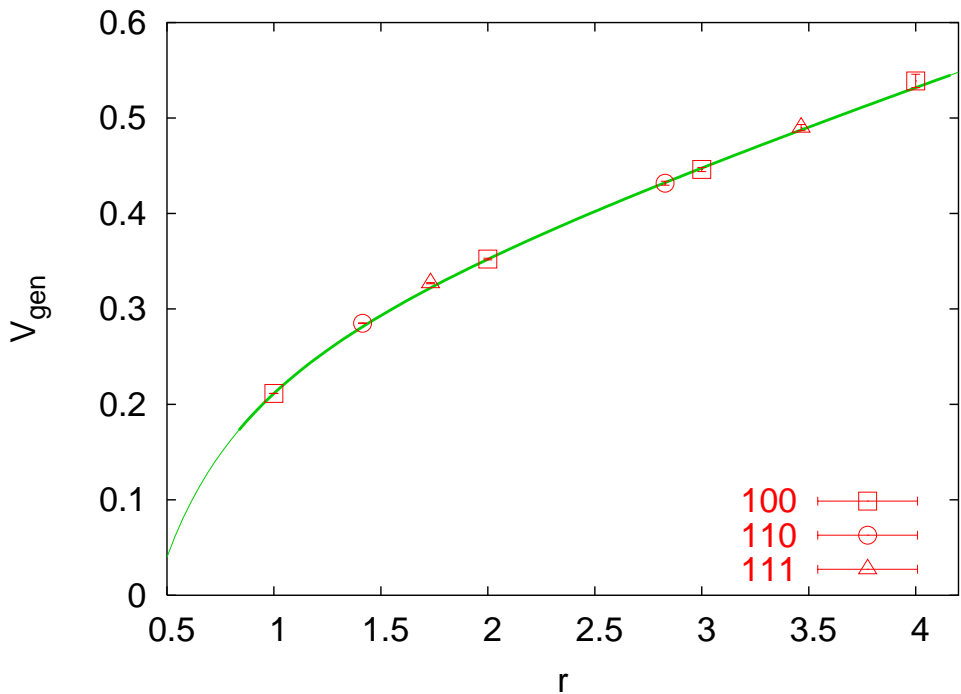}
\caption{\label{fig:pot-off_ext} Static quark-antiquark potential on 
the effective-action-generated  $c=1.50$ lattice; on and 
off axis separations.}
\end{figure}

There is now significant restoration of rotational invariance 
(see Table \ref{tab:ext_act_delta} and Fig. \ref{fig:pot-off_ext}) 
compared to the single plaquette action. 
The fluctuations in $\delta V$ values are
primarily due to the use of the ansatz for $V_{on}$.
\begin{table}[ht]
\centering
\begin{tabular}{|c|c|}
\hline
$c$&$\delta V$\\\hline
0.80&0.0033\\
0.90&0.0024\\
1.00&0.0060\\
1.50&0.0133\\
\hline
\end{tabular}
\caption{\label{tab:ext_act_delta}Amount of rotational 
invariance violation. The same decimations as in the previous table.}
\end{table}

In conclusion, our preliminary results indicate that, provided  
the decimation parameter(s) are 
tuned in the appropriate zone, the extended multi-representation 
(plaquette + rectangle)-action (\ref{eq:ef_act_ext}) does 
 a good job of reproducing under decimation the physics of the 
original system over a wide scale regime.


\acknowledgments
We thank Academic Technology Services (UCLA) 
for computer support. This work was in part supported by 
NSF-PHY-0555693.

\end{document}